\documentclass[11pt]{article}\topmargin=-0.5cm\hfuzz=10pt\oddsidemargin=0.3cm \textheight 220mm \textwidth=16.0cm
\usepackage{amssymb}
\newcommand{\comm}[1]{}

\usepackage{pdfpages}

\def\xxxonly{\rd}
\def\xxxonly{\comm}
\def\short{\comm}

\usepackage{color}
\newcommand{\rd}[1]{\color{red}#1\color{black}} 
\newcommand{\gr}[1]{\color{green}#1\color{black}}

\usepackage{endnotes}

\newtheorem{theorem}{Theorem}
\newtheorem{lemma}{Lemma}
\newtheorem{corollary}{Corollary}
\newtheorem{definition}{Definition}
\newtheorem{remark}{Remark}

\def\e{\varepsilon}

\def\defi{\stackrel{{\scriptscriptstyle \Delta}}{=}}
\def\defi{:=}

\def\d{\delta}
\def\o{\omega}
\def\O{\Omega}

\def\F{{\cal F}}
\def\w{\widehat}
\def\Ind{{\,\rm Ind\,}}
\def\Ind{{\mathbb{I}}}

\def\R{{\bf R}}

\def\PP{{\cal P}}

\def\g{\gamma}
\def\C{{\bf C}}

\def\ww{\widetilde}
\def\X{{\cal X}}
\def\t{\theta}
\def\oo{\bar}

\def\U{{\cal U}}

\newcommand{\be}{\begin{equation}}
\newcommand{\ee}{\end{equation}}
\newcommand{\bd}{\begin{displaymath}}
\newcommand{\ed}{\end{displaymath}}
\newcommand{\ba}{\begin{array}{ll}}
\newcommand{\ea}{\end{array}}
\newcommand{\baa}{\begin{eqnarray}}
\newcommand{\eaa}{\end{eqnarray}}
\newcommand{\baaa}{\begin{eqnarray*}}
\newcommand{\eaaa}{\end{eqnarray*}}
\font\sm=cmr10

\def\PP{{\cal P}}

\def\oo{\bar}

\date{Submitted: July 1, 2022. Revised:  November 20, 2022}
\title{Predictors for high frequency signals based on rational polynomials approximation of periodic exponentials}
\author{Nikolai Dokuchaev\\
{\sm Zhejiang University/University of Illinois at Urbana-Champaign Institute}, 
\\
{\sm  Zhejiang University, Haining, Zhejiang Province, China 314400}}
 
\begin{document}
\def\gr{\comm}
\def\rd{\comm}
 \def\short{\comm}
\def\brea{}
\def\breakk{}
\def\break{}
\def\BRR{}
\def\breakm{\nonumber\\  }\def\BR{}\def\BRR{}
\def\breacm{}
\maketitle
\let\thefootnote\relax\footnote{{To appear in "Problems of Information Transmission"}}
\begin{abstract} The paper presents   linear integral predictors for continuous time  high-frequency signals  with a a finite spectrum gap. The predictors  are based on approximation of a complex valued periodic exponential (complex sinusoid)  
by rational polynomials.  
\end{abstract}

\section{Introduction}
We consider  prediction and predictors for continuous time signals.   A common  approach to forecasting of signals is  based on  removing high-frequency component regarded as a noise and using some filters   and forecasting of smooth low-frequency component, which is presumed to be easier. This approach assumes the loss of the information contained in the high-frequency component that is regarded as a noise.  But there are also works focused the extraction of the information contained  in the high-frequency component.
 These works are based on a variety  of statistical methods and learning models; see, e.g.,
\cite{B,C,G,LH,LT} 
\comm{Brooks and Hinich  (2006), Christensen et al  (2012), Granger (1998), 
Li et al (2020), Luo and Tian   (2020),} and references therein.

We study pathwise predictability and predictors of
continuous time signals in deterministic setting and in the
framework of the frequency analysis. It is well known that certain
restrictions on the  spectrum can ensure 
opportunities for prediction and interpolation of the signals;
see, e.g., \cite{K}-\cite{V} \comm{Knab
(1979), Papoulis (1985), Marvasti (1986), Vaidyanathan (1987), Lyman
{\it et al} (2000, 2001), Lyman and Edmonson (2001), and the bibliography therein.}
These works considered predictability of band-limited signals;
the  predictors obtained therein were non-robust with respect to
small noise in high frequencies; see, e.g., the discussion in
 \comm{Higgins (1996), } \cite{Higgins96}, Chapter 17.
We study predictors for high-frequency
signals,  i.e. for signals without any restrictions on the rate of decay of the spectrum on the higher frequencies.  We consider signals such that their spectrum have a finite  spectral gap, i.e. an interval  where its Fourier transform vanishes. It is known that these signals allow  unique extrapolations from their past observations.  However, feasibility of predicting algorithm is not implied by this uniqueness. In general,
uniqueness  of a path does not ensure possibility to predict this path; some discussion on this can be found in  \cite{D21,D21r} \comm{Dokuchaev (2021a,b).}

Predictors for anticausal convolutions (i.e., integrals including future values) of high frequency signals
were obtained in Dokuchaev (2008), for signals with finite spectral gap,
and in Dokuchaev (2021), for the signals with a  single point spectral degeneracy.\comm{\cite{D08}} \comm{\cite{D21}}
The predictors therein were independent on the
spectral characteristics of  the input signals from a class with certain spectral 
degeneracy for low frequencies. These predictors  depended on the kernels of the corresponding anticausal  convolutions.
 
The present paper offers  some principally new    predictors for the high-frequency signals. The transfer functions for these predictors are polynomials of the inverse $1/\o$ approximating a periodic exponential $e^{i\o T}$, where $\o\in\R$ represents the frequency, and where  $T>0$ represents a  preselected prediction horizon. These predictors allow a compact explicit representation in the time domain and  in the frequency domain. Again, the predictors are independent on the
spectral characteristics of the input signals with fixed and known finite spectral gap.
The  method is based  on the approach from Dokuchaev (2022) for prediction of signals with 
fast-decaying spectrum, where polynomial approximations of the periodic exponential have been used.

The paper is organized in the following manner. In Section
\ref{secDef}, we formulate the definitions and background facts related to
the linear weak predictability. In Section
\ref{secM}, we formulate
 the main theorems on predictability and predictors (Theorem \ref{ThM} and Theorem \ref{Th2}). In Section \ref{secD}, we discuss  some implementation problems.
Section \ref{secProof} contains the proofs.

\section{Problem setting and definitions}\label{secDef}
Let $x(t)$ be a currently observable complex valued continuous time process,
$t\in\R$. The goal is to estimate, at current times $t$, the values
$x(t+T)$, using historical values of the observable process
$x(s)|_{s\le t}$. Here  $T>0$ is a given prediction horizon. \par
 
 We need some notations and definitions.
\par
For $p\in[1,+\infty)$ and a domain  $G\subset \R$, we denote by $L_p(G,\R)$ and $L_p(G,\C)$  the usual $L_p$-spaces
of functions $x:G\to\R$ and $x:G\to \C$ respectively. We denote by $C(G,\R)$ and $C(G,\C)$  the usual linear normed spaces
of bounded continuous functions $x:G\to\R$ and $x:G\to \C$ respectively, with the supremum norm.
\par
For $x\in  L_p(\R,\C)$, $p=1,2$, we denote by $X=\F x$ the function
defined on $i\R$ as the Fourier transform of $x$; $$X(i\o)=(\F
x)(i\o)= \int_{-\infty}^{\infty}e^{-i\o t}x(t)dt,\quad \o\in\R.$$ If
$x\in L_2(\R,\C)$, then   $X(i\cdot)\in L_2(\R,\C)$.
\par
  
Let  $\oo\X$ be the  set of signals $x:\R\to\R$ such that 
their Fourier transforms $X(i\cdot)\in L_1(\R,\C)$.  
In particular, the class $\oo\X$ includes signals formed as $x(t)=\int_t^{t+\d}y(s)ds$  for $y\in L_2(\R,\R)$, $\d\in\R$. Clearly, $\oo\X\subset C(\R,\R)$, i.e., these signals  are bounded and continuous. 

We consider $\oo \X$ as a linear normed space provided with the norm  $\|X(i\cdot)\|_{ L_1(\R,\C)}$, where $X(i\cdot)=\F x$  for $x\in\oo\X$. \gr{(It can be noted that this space is not complete). }

Let $\PP$ be the set of all continuous mappings $p:\oo\X\to C(\R,\C)$ such that, for any $x_1,x_2\in\oo\X$ and $\tau\in\R$, we have that 
 $p(x_1(\cdot))(t)=p(x_2(\cdot))(t)$ for all $t\le\tau$   if $x_1(t)=x_2(t)$ for all $t\le \tau$. In other words, this is the set of "causal" mappings;
we will look for predictors in  this class.

Let $T>0$ be given.

\begin{definition} Let $\X\subset\oo\X$.
\begin{itemize} 
\item[(i)]
 We say that the class $\X$ is linearly predictable 
 with the prediction horizon $T$ if there exists a sequence $\{\ww
p_{d}(\cdot)\}_{d=1}^{+\infty}\subset\PP$ such that $$
\sup_{t\in\R}|x(t+T)-\ww y_{d}(t)|\to 0\quad \hbox{as}\quad
d\to+\infty\quad\forall x\in\X, $$ where \baaa &&
\ww y_{d}= \ww p_d(x(\cdot)).\label{predict} \eaaa 
\item[(ii)] 
 We say that the class $\X$ is  uniformly linearly predictable 
  with the prediction horizon $T$  if   there exists a sequence $\{\ww
p_{d}(\cdot)\}_{d=1}^{+\infty}\subset\PP$  such that \baaa \sup_{t\in\R}|x(t+T)-\ww
y_{d}(t)|\to 0\quad\hbox{uniformly in} \quad x\in\X,\eaaa 
where  $\ww
y_{d}(\cdot)$ is as in part (i) above.
\end{itemize}
\end{definition}

Functions $\ww y_{d}(t)$ in the definition above
can be considered as approximate predictions of the process $x(t+T)$.

\section{The main result} 
\label{secM}
\def\RR{\cal R}
\def\PP{\cal P}

Let $\O>0$ be given, and let 
$\X_\O$   be the set of all signals
$x(\cdot)\in\oo\X$ such that 
$X(i\o)=0$ for $\o\in(-\O,\O)$, for  $X=\F x$.

Let $\U_\O$ be some  set of 
signals $x\in \X_\O$ such that  $\|X(i\cdot) \|_{L_{1}(\R,\C)}\le 1$
and
 $\int_{\o:\ |\o|\ge M}|X(i\o)|d\o\to 0$ as $M\to +\infty$ uniformly over $x\in\U_\O$, for  $X=\F x$.

\vspace{0.4cm}  
\begin{theorem}\label{ThM}  For any  $T>0$, the following holds.  
\begin{itemize}\item[(i)] The class $\X_\O$ is linearly predictable  with the prediction horizon $T$.
\item[(ii)] 
The class $\U_\O$ is uniformly linearly predictable  with the prediction horizon $T$.
\end{itemize}
\end{theorem}
\subsection{A family of predictors}

In this section, we introduce some predictors. 
 \par

 For $d=1,2,...$, let $\Psi_d$ be the set of all functions $\sum_{k=1}^d \frac{a_k}{z^k}$ defined for $z\in\C\setminus \{0\}$, for all $a_k\in\R$.  Let $\Psi\defi \cup_{d}\Psi_d$.
 
 For $d=0,1,2,...,$ and $s\in\R$, let  $\X^{(d)}(s)$ be the set of all signals $x\in \oo\X$ such that $\int_{-\infty}^s|t^{d}x(t)|dt<+\infty$. It can be noted that this class includes, in particular, signals 
 $x\in\oo\X$ such that $\int_\R\left|\frac{d^k X}{d\o^k}(i\o)\right|^2d\o<+\infty$ for $k=0,1,...,d+1$, $X=\F x$.

Let  $r:\R\to (0,1]$ be a continuous function such that $r(0)=1$, that
 $r(\o)\equiv r(-\o)$,  that the function $r(\o)$ is monotonically non-increasing on $(0,+\infty)$, and that 
  $r(\o)\to 0$ as $|\o|\to +\infty$. Let $r_\nu(\o)\defi r(\nu\o)$, $\nu\in(0,1]$. 
\vspace{0.4cm}

\begin{theorem}\label{Th2} 
\begin{itemize}
 \item[(i)]
For any   $\e_1>0$ and any $x\in\X_\O$ such that $\|X(i\cdot) \|_{L_{1}(\R,\C)}\le 1$, there exists $\nu_0=\nu_0(\e_1,x)>0$ such that, for $X=\F x$ and any $\nu\in (0,\nu_0]$, 
\baa
&&\int_{\o:\ |\o|\ge \O} (1-r_\nu(\o))|X(i\o)|d\o\le\e_1.\label{e1}
 \eaa
 Moreover, one can select  the same $\nu_0=\nu_0(\e_1)$ for all $x\in \U_\O$. 
\item[(ii)]
For any  $\e_2>0$ and  $\nu>0$,  there exists  a integer $d=d(\nu,\e_2,T)>0$ and $\psi_d\in\Psi_d$ such that 
\baa
\sup_{\o:\ |\o|\ge \O} 
|e^{i\o T}r_\nu(\o)-\psi_d(i\o)|  \le \e_2. \label{e2}
\eaa
\item[(iii)] 
The   predictability considered in Theorem \ref{ThM}(i)
for   $x\in \X_\O$, as well as the   predictability considered in Theorem \ref{ThM}(ii)
for   $x\in \U_\O$,
 can be ensured with
 the sequence of the predictors $ p_d:\X_\O\to C(\R,\C)$, $d=1,2,....,$ defined by their transfer functions $\psi_d (i\o)$. More precisely, for any $\e>0$ and  $\w y_d(t)= p_d(x(\cdot))(t)$, the estimate  
\baaa
\sup_{t\in\R}|x(t+T)-\w y_d(t)|\le\e
\eaaa
holds if   $\nu$, $d$, and $\psi_d$ are such that
(\ref{e1})-(\ref{e2}) hold for sufficiently small $\e_1$ and $\e_2$  such that
\baaa
\e_1+\e_2\le 2\pi\e.
\eaaa
It can be noted that, for the inputs $x\in \X_\O$, the  transfer functions  $\psi_d (i\o)$ can be replaced by the functions $\psi_d (i\o) \Ind_{\o: |\o|\ge \O}$, where $\Ind$ denotes the indicator function.

 \item[(iv)]  
 For  $x\in  \X^{(d-1)}(t)\cap\X_\O$, the predictors described above can be represented  as 
 \baa
  p_d(x(\cdot))(t)=\int_{-\infty}^t K(t-\tau)x(\tau)d\tau,
  \label{Kk}
\label{pred} \eaa
where 
\baaa
K(t)=\sum_{k=1}^d a_k \frac{t^{k-1}}{(k-1)!}.
 \eaaa
  Here
 $a_k\in\C$ are the coefficients for $\psi_d(z)=\sum_{k=1}^d a_k z^{-k}$ from part (ii).
\end{itemize}
\end{theorem}
\subsection{Integral representation of the predictors for general type $x\in\X_\O$}
\label{secIR}
Representation (\ref{Kk}) for the predictors  above requires that  $x\in  \X^{(d-1)}(t)$.
Let us discuss possibilities of representations in time domain for general type $x\in\X_\O$.

Consider operators $h_k$ defined on $\X_\O$  by their transfer functions 
$(i\o)^{-k}$, $k=1,2,...$\,. In other words, if $y=h_k(x)$ for $x\in\X_\O$, then $Y(i\o)=(i\o)^{-k}X(i\o)$ for  $Y=\F y$ and $X=\F x$. Clearly, $h_{k}(x(\cdot))\in\X_\O$, the Fourier transforms of processes
$h_{k}(x(\cdot))$ vanish on $[-\O,\O]$,   and the operators 
$h_k:\X_\O\to C(\R,\C)$ are continuous.    By the definitions, it follows that \baaa
  p_d(x(\cdot))(t)=\sum_{k=1}^d a_k h_k(x(\cdot))(t). \label{Kkk}
\label{pred2} \eaaa
It can be noted that   $p_d(\cdot)$ depends on  $T$ via the coefficients $a_k$ 
 defined for $\psi_d(\o)$ approximating  
 $e^{i\o T}$.

Formally, the operator $h_k(x(\cdot)$ can be represented as 
\baa
\hspace{-0.5cm}&&h_k(x(\cdot))(s_k)\breakk= \int_{-\infty}^{s_k}\!\!\! ds_{k-1}\int_{-\infty}^{s_{k-1}}\!\!\! ds_{k-2}.... \int_{-\infty}^{s_2}\!\!\! ds_1
\int_{-\infty}^{s_1} x(s)ds, \quad
\label{int1}\eaa
i.e.,
\baaa
&&h_1(x(\cdot))(t)=\int_{-\infty}^t x(s)ds,\quad \breakk h_{k}(x(\cdot))(t)=\int_{-\infty}^t (h_{k-1}(x(\cdot))(s)ds, \quad k=2,3,...
\label{hint}\eaaa
\par
For general  type   $x\in\X_\Omega$, there is no guarantee that  $x\in L_1(\R,\R)$
or $ h_k(x(\cdot)) \in L_1(\R,\R)$. However, the integrals above are well defined, because
they can be replaced by  integrals over finite time intervals
\baa
&&h_1(x(\cdot))(t)=\int_{R_1}^t x(s)ds,\quad 
 \breakk h_{k}(x(\cdot))(t)=\int_{R_k}^t (h_{k-1}(x(\cdot))(s)ds, \quad   k>1,
\label{int2}\eaa
where $R_k$ are roots of signals $h_{k}(x(\cdot))(t)$. This is possible because of the special properties of real signals with Fourier transform vanishing on an interval: for any $\tau<0$, these signals have infinitely many roots in the interval $(-\infty,\tau)$; see, e.g., \cite{BU}.  
Hence the predictors defined in Theorem \ref{Th2} for $x\in\X_\O$  allow  an alternative integral representation via (\ref{int1}) or (\ref{int2}). \short{\vspace{0.4cm}  }
 \section{ On numerical implementation of the predictors} \label{secD}
The direct implementation of the predictor introduced in Theorem \ref{Th2}  requires 
evaluation of integrals over semi-infinite intervals that could be numerically challenging.  
However, this theorem could lead to predicting methods  bypassing  this calculation.
Let us discuss these possibilities.
\par
 
Let  $t_1\in\R$ be given. Let $x_k\defi h_k(x)$ for  $x\in\X_\O$,  $k=1,2,..$,  and let $\eta_k\defi x_k(t_1)$.

\begin{lemma}\label{lemma1}  In the notation of Theorem \ref{Th2}, for any $t\ge t_1$, we have  that   $\w y_d=p_d(x(\cdot))$ can be represented as
\baa
\w y_d(t)=
\sum_{k=1}^d a_k \left(\sum_{l=1}^k c_{l}(t)\eta_l+f_{k}(t)\right),
\label{viaeta0}\eaa
where $c_{l}(t)\defi (t-t_1)^{(l-1)}/(l-1)!$ and
\baaa
f_{k}(t)\defi  \int_{t_1}^{t}\! d\tau_1\int_{t_1}^{\tau_1} \! d\tau_2...\!\int_{t_1}^{\tau_k} x_{}(s)ds.
\eaaa 
\end{lemma}
This lemma shows that calculation of the prediction $\w y_d(t)$  of $x(t+T)$ is easy for $t>t_1$  if we know all $\eta_k$ and observe $x|_{[t_1,t]}$.

Let us discuss possible ways  to evaluate $\eta_k$ bypassing direct integration over infinite intervals. 

First, let us observe that (\ref{viaeta0})  implies a useful property given below.
\begin{corollary}\label{corr1} For any $\e>0$, there exist
 an integer $d=d(\e)>0$ and $a_1,....,a_d\in\R$  such that,
for any $x\in\X_\O$ and  $t_1\in \R$,
there exist  $\oo\eta_1,....,\oo\eta_d\in\R$  such that 
 $|x(t+T)-y_d(t)|\le \e$  for all $t\ge t_1$, for \baa
 y_d(t)=y_d(t,\oo\eta_1,...,\oo\eta_d)\defi
 \sum_{k=1}^d a_k \left(\sum_{l=1}^k c_{l}(t)\oo\eta_l+f_{k}(t)\right).
\label{viaeta}\eaa
\end{corollary}
\vspace{1mm}
In this corollary, $d=d(\e)$ can be selected  as defined  in Theorem \ref{Th2}(i)-(ii), where $\e_1$ and $\e_2$
are such that $\e_1+\e_2\le 2\pi \e$.
\par
Further, let us discuss using  (\ref{viaeta0}) for evaluation $\eta_k$ and prediction. 
Let $\t>t_1$. Assume that  the goal is to forecast the value $x(\t\!+\!T)$ 
 given observations at times $t\le  \t$.  
It appears that if  $\t>t_1+T$ then Corollary \ref{corr1} gives an opportunity to construct predictors  via fitting  parameters  $\eta_1,...,\eta_d$ using past observations available for $t\in [t_1,\t-T]$: we  can match  the values  $y_d(t,\oo\eta_1,...,\oo\eta_d)$ with the
 past observations $x(t+T)$. Starting from now, we assume that $\t>t_1+T$.

Let $d$ be large enough  such that
 $x(t+T)$ is approximated  by $\w y_d(t)$ as described in Theorem \ref{Th2}, i.e.,
 $\sup_{t\in\R}|x(t+T)-\w y_d(t)|\le \e$  for some sufficiently small 
   $\e>0$, for some  choice of  $a_k$.

 As an approximation of the true $\eta_1,...,\eta_d$, 
we can accept a set $\oo\eta_1,...,\oo\eta_d$ such that
 \baa
|x(t+T)-y_d(t,\oo\eta_1,...,\oo\eta_d)|\le \e\quad \forall t\in [t_1, \t-T].
 \label{eta1}\eaa
 (Remind that, at time $\t$,  values  $x(t+T)$ and $y_d(t,\oo\eta_1,...,\oo\eta_d)$
 are observable  for these $t\in[ t_1,\t-T]$).  If (\ref{eta1}) holds, we can conclude that
  $y_d(t,\oo\eta_1,...,\oo\eta_d)$ delivers an acceptable prediction of $x(t+T)$ for these $t$.
 Clearly, Theorem \ref{Th2} implies that a set $\oo\eta_1,...,\oo\eta_d$ ensuring (\ref{eta1})  
  exists since this inequality holds with $\oo\eta_k=\eta_k$.
 
 The corresponding value  $y_d(\t,\oo\eta_1,...,\oo\eta_d)$ would give an estimate for $\w y_d(\t)$ and, respectively, for $x(\t+T)$.

Furthermore,  finding a set $\oo\eta_1,...,\oo\eta_d$ ensuring (\ref{eta1})  could still be difficult.  Instead, one can consider fitting predictions and observations at a finite  number of points $t\in[t_1,T-\t]$.

Let  a integer $\oo d\ge d$ and  a set $\{t_m\}_{m=1}^{\oo d}\subset \R$ be selected such that $t_1<t_2<t_3<...<t_{\oo d-1}<t_{\oo d}\le\t-T$.  We suggest to  use observations  $x(t)$ at times $t=t_m$. 
Consider  a system of equations
\baa
\sum_{k=1}^d a_k \left( \sum_{l=1}^k c_{l}(t_m)\oo\eta_l+f_{k}(t_m) \right)=\zeta_m, \quad m=1,...,\oo d.\label{sys3} \eaa
\par
Consider first the case where $\oo d=d$. In this case, we can select $\zeta_m= x(t_m+T)$; these values are directly observable (without calculation of integrals of semi-infinite intervals required for $\w y_d(t_m)$).  The corresponding choice of $\oo\eta_k$ ensures zero prediction error for $x(t_m+T)$, $m=1,...,\oo d$. 

Including into consideration more observations, i.e.,  selecting larger $\oo d>d$ and wider interval $[t_1,\t-T]$, would
 improve estimation of $\eta_k$.
  If we consider $\oo d>d$, then, in the general case,  it would not be feasible to achieve that 
  $y_d(t,\oo\eta_1,...,\oo\eta_d)= x(t_m+T)$ for all $m$, since it cannot be guaranteed  that   system (\ref{sys3})
  is solvable for   $\zeta_m\equiv  x(t_m+T)$: the system will be overdefined.  
  Nevertheless, estimate presented in  (\ref{eta1})  can still be achieved for any arbitrarily  large $\oo d$, since  (\ref{eta1}) holds.  A solution  could be found using  
methods for fitting linear models.

Furthermore, instead of calculation of the coefficients $a_k$ via solution of the approximation problem for the complex exponential described  in Theorem \ref{Th2}(i)-(ii), one may  find these coefficients considering them  as
additional unknowns  in system (\ref{sys3}) with $\oo d\ge 2d$.
Theorem \ref{Th2} implies again that there exist $\oo\eta_k=\eta_k\in\R$ and $a_k\in \R$ such that (\ref{sys3}) holds with $\zeta_m=\w y_d(t_m)$. This would lead to a 
nonlinear fitting problem for unknowns $a_1,...,a_d,\oo\eta_1,...,\oo\eta_d$.

 So far,  the consistency of these estimates is unclear since  a choice of smaller $\e$ leads to larger $d$. We leave analysis of these methods for the future research. 
\xxxonly{
 \begin{remark}  It appears that an existence of a consistent method of recovery  of all $\eta_k$ from  observations on the time interval $(t_1,\t-T)$, as described above, would imply that high-frequency signals have unique extrapolation from finite time interval, similarly to the band-limited signals.  We leave analysis of this for the future research as well. 
 \end{remark}}
 \section{Proofs}\label{secProof}
Theorem \ref{ThM} follows immediately from Theorem \ref{Th2}. \short{ Let us prove Theorem \ref{Th2}.}
\par
{\em Proof of Theorem \ref{Th2}}.  
Let us prove statement (i).  Let us select $M>0$ such that 
\baaa
\int_{\o:\ |\o|>M} |X(i\o)|d\o<\e_1/2 \quad \forall x\in\U_\O. 
\eaaa
We have that $r_\nu(\o)\to 1$ uniformly in $\o\in[-M,M]$, since $0<1-r_\nu(\o)\le 1-r_\nu(M)$ for $\o\in[-M,M]$. Hence one can select $\nu>0$ such that 
$\int_{-M}^M (1- r_\nu(\o))|X(i\o)|d\o\le \e_1/2$. This implies that
\baaa
\int_{-\infty}^\infty (1- r_\nu(\o))|X(i\o)|d\o\brea=
 \int_{-M}^M (1- r_\nu(\o))|X(i\o)|d\o+\int_{\o:\, |\o|>M} (1- r_\nu(\o))|X(i\o)|d\o\\
\le \int_{-M}^M (1- r_\nu(\o))|X(i\o)|d\o+\int _{\o:\,|\o|>M}|X(i\o)|d\o
\brea \le \frac{\e_1}{2}+\frac{\e_1}{2}\le\e_1.
\eaaa
This completes the proof of statement (i). 
  
Let us prove statement (ii).  By  the 
Stone-Weierstrass theorem for real continuous functions on locally compact spaces, it follows 
 that  there exist $\psi_d^c(\o)\in \Psi_d$  and $\psi_d^s(\o)\in \Psi_d$  such that
\baaa
&& \sup_{\o: |\o|\ge \O} |\cos(T\cdot) r_\nu(\cdot)-\psi_d^c(\cdot)|\le\e_2/2, \quad\breakk 
\sup_{\o: |\o|\ge \O} |\sin(T\cdot) r_\nu(\cdot)-\psi_d^s(\cdot)|\le \e_2/2.
\eaaa
(See, e.g.,  Theorem 12 in \cite{SW}, pp. 240-241). 

It is easy to see that it suffices to select odd functions 
$\psi_d^c(\o)=\sum_{k=1}^d \g_k^c \o^{-k}$ and even functions 
$\psi_d^s(\o)=\sum_{k=1}^d \g_k^s \o^{-k}$, i.e., $\g_{2m+1}^c=0$  and 
$\g_{2m}^s=0$ for integers $m\ge 0$. Here $\g_k^c$ and $\g_k^s$  are real. 

We construct the desired functions as 
\baaa
\psi_d(i\o)=
\psi_d^c(\o)+i\psi_d^s(\o)=\sum_{k=1}^d \g^c_k \o^{-k}+i\sum_{k=1}^d \g_k^s \o^{-k}=\sum_{k=1}^d a_k (i\o)^{-k },\eaaa
where  the coefficients $a_k\in\R$ are defined as the following:

\begin{itemize}
\item
If $k=2m$ for an integer $m$, then
$a_k=(-1)^m \g^c_k$. 
\item
If $k=2m+1$ for an integer $m$, then
$a_k=-(-1)^m \g^s_k$. 
\end{itemize} 
This choice of $\psi_d$  ensures that estimate (\ref{e2}) holds. This completes the proof of statement (ii)

Let us prove statement (iii).  Assume that estimates (\ref{e1})-(\ref{e2}) hold for selected $d,\nu,\psi_d$. We have that 
\baaa
&&2\pi(x(t+T)-\w y_d(t))=\int_{-\infty}^\infty e^{i \o t} (e^{i\o T}-\psi_d(i\o) )X(i\o)d\o =
A(t)+B(t),
\eaaa
where 
\baaa
A(t)=\int_{-\infty}^\infty e^{i \o t} (e^{i\o T}- e^{i\o T}r_\nu(\o))X(i\o)d\o
, \\
B(t)=\int_{-\infty}^\infty e^{i \o t} (e^{ i\o T}r_\nu(\o)-\psi_d(i\o) )X(i\o)d\o.
\eaaa
Clearly, 
\baaa
|A(t)|\le \int_{-\infty}^\infty (1- r_\nu(\o))|X(i\o)|d\o\le \e_1
\eaaa and \baaa
|B(t)|&\le& \int_{-\infty}^\infty |e^{ i\o T}r_\nu(\o)-\psi_d(i\o)| |X(i\o)| d\o
\\ &\le& 
\sup_{\o:\ |\o|\ge \O} |e^{ i\o T}r_\nu(\o)-\psi_d(i\o) |\int_{-\infty}^\infty|X(i\o)| d\o\le \e_2.
\eaaa
Hence $2\pi|x(t+T)-\w y_d(t)|\le \e_1+\e_2$.  This proves the uniform predictability considered in Theorem \ref{ThM}(ii)
for signals  $x\in \U_\O$. 
The predictability considered in Theorem \ref{ThM}(i) follows immediately from the proof above applied to singletons  $\U_\O=\{x(\cdot)\}$, multiplied on a constant, if needed, to bypass the restriction that 
$\|X(\cdot)\|_{L_1(\R,\C)\le 1}$.
This completes the proof of statement (iii).
  
Let us prove statement (iv). First, 
the known properties of Fourier transforms of derivatives and antiderivatives imply representations (\ref{int1})-(\ref{int2}); see  some clarifications  in Section  \ref{secIR}.
The statement (iv) can be obtained by the consequent application of the Fubini's Theorem to integrable in $L_1((-\infty,t],\R)$ signals
$(\tau-s)^\ell x(s)$ presented in (\ref{int1}) for $\ell=1,2,...$, $\tau\in (\infty,s]$.

{\em Proof of Lemma \ref{lemma1}}.
In the notation of Theorem \ref{Th2},  we have  that   
 $y_d(t)=\sum_{k=1}^d a_k x_k(t)$ for any $t\ge t_1$, i.e.,
  \baa
y_d(t) =\sum_{k=1}^d a_k \left(\eta_k+\int_{t_1}^{t}x_{k-1}(s)ds\right).
 \label{sys} \eaa
(Here we assume that $x_0\defi x$).
Further, we have  that  
 \baaa
 \int_{t_1}^{t_{}}x_{1}(t)dt=\int_{t_1}^{t_{}}\left(\eta_1 +\int_{t_1}^\tau x_{0}(s)ds\right)d\tau\brea=\eta_1(t_{}-t_1)+\int_{t_1}^{t_{}}  d\tau 
 \int_{t_1}^\tau x_{}(s)ds
 \eaaa
 and
 \baaa 
\int_{t_1}^{t_{}}x_{2}(t)dt=\int_{t_1}^{t_{}}\left(\eta_2 +\int_{t_1}^{\tau_1} x_{1}(s)ds\right)d\tau_1\brea=\eta_2(t_{}-t_1)+\int_{t_1}^{t_{}}  d\tau_1 
 \int_{t_1}^{\tau_1} x_{1}(s)ds\\ =
 \eta_2(t_{}-t_1)+\int_{t_1}^{t_{}}  d\tau_1 
\left[\eta_1(\tau_1-t_1)+\int_{t_1}^{\tau_1} d\tau_2 \int_{t_1}^{\tau_2}x_{}(s)ds\right]
\\=
 \eta_2(t_{}-t_1)+\frac{\eta_1^2}{2}(t_{}-t_1)
+\int_{t_1}^{t_{}}  d\tau _1\int_{t_1}^{\tau_2} x_{}(s)ds.
 \eaaa
Similarly, we obtain that
\baaa
&& \int_{t_1}^{t_{}}x_{k}(t)dt= \eta_k(t_{}-t_1)+\frac{\eta_{k-1}}{2}(t_{}-t_1)^2+...\breakk +
 \frac{\eta_{1}}{k!}(t_{}-t_1)^k
+\int_{t_1}^{t_{}} d\tau_1\int_{t_1}^{\tau_1} d\tau_2...\int_{t_1}^{\tau_k} x_{}(s)ds.
 \label{etak}\eaaa
\par
It follows that
\baaa\eta_k+\int_{t}^{t}x_{k-1}(s)ds=
\sum_{l=1}^k c_{l}(t)\eta_l+f_{k}(t).
\label{eta22}\eaaa
Together with (\ref{sys}), this proves (\ref{viaeta}) and completes the proof of Lemma \ref{lemma1}. $\Box$

\end{document}